\documentclass[runningheads]{llncs}

\usepackage{iciapabbrv}

\usepackage{graphicx}
\usepackage{booktabs}

\usepackage[accsupp]{axessibility}  

\usepackage{hyperref}

\usepackage{orcidlink}

\begin{document}

\title{Mapping Emotions in the Brain: A Bi-Hemispheric Neural Model with Explainable Deep Learning} 

\titlerunning{Mapping Emotions in the Brain}

\author{David Freire-Obregón\inst{1}\orcidlink{0000-0003-2378-4277} \and
Agnieszka Dubiel\inst{2}\orcidlink{0000-0001-8758-7593} \and
Prasoon Kumar Vinodkumar\inst{3}\orcidlink{0000-0002-9741-5848}
\and
Gholamreza Anbarjafari\inst{3,4,5}\orcidlink{0000-0001-8460-5717} \and
Dorota Kamińska\inst{2}\orcidlink{0000-0002-3416-5554}
\and
Modesto Castrillón-Santana\inst{1}\orcidlink{0000-0002-8673-2725}
}

\authorrunning{D. Freire-Obregón et al.}

\institute{SIANI, Universidad de Las Palmas de Gran Canaria, Spain\\
\email{david.freire@ulpgc.es} \and
Institute of Mechatronics and Information Systems, Lodz University of Technology, Poland \and 
3S Holding, Tartu 51011, Estonia \and
PwC Finland, Itämerentori 2, 00180 Helsinki, Finland \and Estonian Business School, A. Lauteri tn 3, 10114 Tallinn, Estonia
}
\maketitle

\begin{abstract}
Recent advances have shown promise in emotion recognition from electroencephalogram (EEG) signals by employing bi-hemispheric neural architectures that incorporate neuroscientific priors into deep learning models. However, interpretability remains a significant limitation for their application in sensitive fields such as affective computing and cognitive modeling. In this work, we introduce a post-hoc interpretability framework tailored to dual-stream EEG classifiers, extending the Local Interpretable Model-Agnostic Explanations (LIME) approach to accommodate structured, bi-hemispheric inputs. Our method adapts LIME to handle structured two-branch inputs corresponding to left and right-hemisphere EEG channel groups. It decomposes prediction relevance into per-channel contributions across hemispheres and emotional classes. We apply this framework to a previously validated dual-branch recurrent neural network trained on EmoNeuroDB, a dataset of EEG recordings captured during a VR-based emotion elicitation task. The resulting explanations reveal emotion-specific hemispheric activation patterns consistent with known neurophysiological phenomena, such as frontal lateralization in joy and posterior asymmetry in sadness. Furthermore, we aggregate local explanations across samples to derive global channel importance profiles, enabling a neurophysiologically grounded interpretation of the model’s decisions. Correlation analysis between symmetric electrodes further highlights the model’s emotion-dependent lateralization behavior, supporting the functional asymmetries reported in affective neuroscience.
\keywords{EEG \and Emotion Recognition \and Explainable AI \and LIME \and Hemispheric Specialization}
\end{abstract}

\section{Introduction}
\label{sec:intro}

Emotion plays a central role in human behavior, influencing decision-making, communication, and social interaction \cite{ledoux2018subjective,Salas-Caceres2024}. Accurately recognizing emotional states is essential for the development of affective computing systems, brain-computer interfaces, and emotion-aware agents \cite{alarcao2017emotions}. Among the various modalities for emotion recognition, electroencephalography (EEG) provides a non-invasive and direct measure of brain activity with high temporal resolution, making it a promising source of information for decoding internal affective states.

Recent work has demonstrated that neural architectures inspired by brain organization, particularly hemispheric specialization, can enhance EEG-based emotion recognition. In this context, bi-hemispheric models, which process signals from the left and right hemispheres independently, have shown improved performance by capturing asymmetrical neural activation patterns associated with emotional processing  \cite{9105104}. A dual-stream recurrent neural network architecture \cite{FreireFG24}, designed to reflect hemispheric specialization, was proposed and evaluated in the FG2024 Brain Responses to Emotional Avatars Challenge \cite{DubielFG24}. This model demonstrated competitive performance on the EmoNeuroDB dataset, which comprises EEG recordings collected during immersive virtual reality scenarios involving emotionally expressive avatars.

Despite these advances, a critical barrier remains: the lack of interpretability in deep learning models applied to neurophysiological data. Most high-performing models operate as black boxes, providing limited insight into the decision-making process and hindering their adoption in clinical, cognitive, and ethically sensitive applications. Understanding why a model associates specific EEG patterns with emotions, as well as the brain regions and temporal segments involved in that inference, is essential for aligning machine learning decisions with neuroscientific knowledge.

To address this gap, we propose an explainability framework that utilizes Local Interpretable Model-Agnostic Explanations (LIME) in conjunction with a bi-hemispheric EEG model \cite{LIME16}. The method adapts LIME to structured dual-input architectures, enabling the quantification of channel-wise importance across hemispheres and emotional categories. Through this approach, model predictions are aligned with neurophysiological plausibility, offering interpretable, localized, and emotion-specific insights into how deep models process affective brain signals. The objective of this work is three-fold: (i) to identify which EEG features and channels contribute most significantly to emotion classification, (ii) to visualize these relevance patterns by mapping them onto brain cortical maps for each emotional category, and (iii) to analyze cross-emotion relevance profiles, revealing potential shared or divergent neural substrates among emotions.

\section{Related Work}
\label{sec:relatedwork}

\textbf{EEG-Based Emotion Recognition}. EEG enables non-invasive, temporally precise monitoring of brain activity, making it well-suited for emotion recognition in domains such as healthcare, neuromarketing, and human-computer interaction (HCI) \cite{Britton06}. Despite this promise, emotion classification from EEG remains difficult due to the signals’ nonlinear and high-dimensional nature.
Emotion modeling often follows either a discrete (e.g., joy, anger) or dimensional (e.g., valence, arousal) approach, with the latter being more prevalent in EEG studies \cite{Matlovic16}. Standard pipelines include preprocessing, feature extraction (e.g., spectral power, connectivity measures), and classification.
Classifiers range from topology-invariant methods, such as Support Vector Machines (SVMs), to topology-aware models, including Convolutional Neural Networks (CNNs) and Graph Neural Networks (GNNs) \cite{Egger19}. However, many struggle with long-range dependencies, which recent dual-stream RNN architectures aim to address by separately processing hemispheric signals with spatially-informed designs \cite{Nianyin18}.

\textbf{Hemispheric Specialization in Affective Neuroscience}. The integration of deep learning methods such as Recurrent Neural Networks (RNNs), CNNs, and GCNs with classical signal processing techniques has significantly improved classification performance in emotion recognition tasks \cite{Freire09,Freire23}. Moreover, the adoption of transfer learning and domain adaptation has enhanced the generalizability of models across different individuals and experimental setups \cite{Gao23}.

Multimodal approaches have also emerged as a promising direction. By combining EEG with complementary signals, such as electrocardiogram (ECG) or facial expressions, studies have improved the robustness and ecological validity of emotion classification \cite{Jenke14,Li17}. For instance, the DREAMER dataset integrates EEG and ECG using affordable consumer-grade equipment to support emotion detection in real-world settings. However, it introduces trade-offs in data quality and generalization \cite{Li17}.


\textbf{Explainable AI for EEG}. Finally, several surveys have systematized the methodological landscape of EEG-based affective computing. For example, some studies highlight the diversity of approaches in stimuli presentation, emotion elicitation, and classification strategies, advocating for multimodal integration and deeper neurophysiological interpretability \cite{Pantic03}. These overviews underscore the field’s shift from purely performance-driven goals toward models that are also explainable and aligned with neuroscientific knowledge.

In this context, our work builds upon the emerging trend of incorporating domain priors, such as hemispheric specialization, into the model design and complements it with post-hoc interpretability techniques. Specifically, we adapt the LIME framework to dual-branch RNN architectures, aiming to bridge the gap between high-performing black-box models and the need for interpretable, neurophysiologically grounded emotion decoding.

\begin{figure*}[t]  
    \centering
    \includegraphics[scale=0.50]{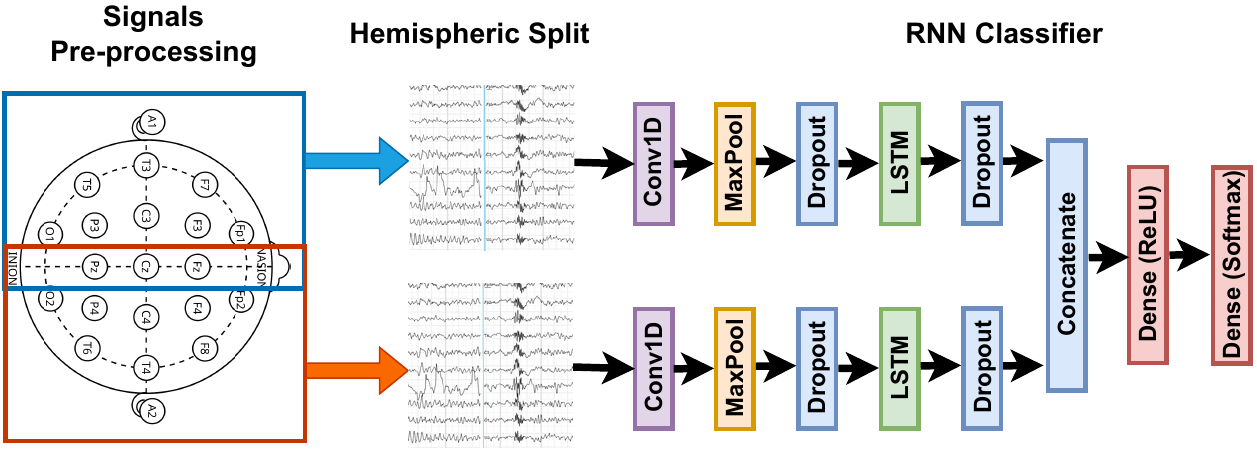}
    \caption{Diagram illustrating the three-module stages for bi-hemispheric emotion recognition from EEG signals: Signal Pre-processing, Hemispheric Split, and Signal Classification using RNN. Adapted from \cite{FreireFG24}.}
     \label{fig:classifier}
\end{figure*}

\section{Methodology}
Our methodological pipeline consists of two main components: a dual-stream neural architecture designed to exploit hemispheric asymmetries in EEG data and a post-hoc interpretability framework based on LIME adapted to such structured input. We first describe the bi-hemispheric classifier used for emotion recognition, which serves as the backbone for our interpretability analysis. Then, we detail how LIME was extended to operate on this architecture, allowing us to extract channel-level relevance scores for each emotional category. 

\subsection{Bi-Hemispheric Emotion Classifier}
\label{sec:bihemispheric-model}

To evaluate our interpretability framework, we apply LIME to the bi-hemispheric neural model for EEG-based emotion recognition \cite{FreireFG24}. According to the authors, this model was designed to leverage hemispheric specialization by processing left- and right-brain signals through parallel but independent branches (see Figure \ref{fig:classifier}). Each branch receives frequency-domain EEG features extracted from one hemisphere, allowing the network to capture lateralized patterns of emotional processing.

\textbf{Input Representation.} The EEG signals are first re-referenced to mastoid electrodes, band-pass filtered, and corrected for filter-induced delay. Then, the signals are transformed into the frequency domain using the Fast Fourier Transform (FFT), resulting in magnitude-based spectral representations. Electrodes are grouped into left and right hemispheres based on standard EEG montages, forming two structurally distinct inputs to the model.

\textbf{Dual-Branch Architecture.} As depicted in Figure \ref{fig:classifier}, the architecture consists of two symmetrical yet non-weight-sharing streams, each responsible for processing one hemisphere’s input. Within each stream, a 1D convolutional layer captures local spectral patterns, followed by max pooling and dropout for regularization. The resulting features are reshaped and passed through a Long Short-Term Memory (LSTM) layer, enabling the model to learn temporal dependencies in the EEG signal dynamics. Another dropout layer is applied after the LSTM to mitigate overfitting.

\textbf{Fusion and Classification.} Outputs from both hemispheric branches are flattened and concatenated, producing a joint representation that is then passed through a dense layer with ReLU activation and L2 regularization. A softmax output layer maps the representation to six emotion classes (e.g., joy, sadness, fear, anger, surprise, and disgust) optimized using categorical cross-entropy loss and the Adam optimizer.

\textbf{Challenge Performance.}  
The bi-hemispheric model achieved an average validation accuracy of 22.78\% on the competition leaderboard, securing the second-highest score among all submitted solutions. Notably, it outperformed all provided baselines, including SVM (17.78\%), LightGBM (19.44\%), and Random Forests (19.44\%). In class-wise performance, this model demonstrated strong recognition of emotions such as joy (43.33\%) and anger (33.33\%), while maintaining balanced performance across other categories. These results underscore the effectiveness of incorporating hemispheric specialization into EEG-based emotion recognition, providing a strong foundation for our subsequent interpretability analysis. In this work, rather than modifying the model architecture or training process, we focus on augmenting it with post-hoc interpretability through LIME in order to understand the specific contributions of channels, frequencies, and hemispheres to the final emotion predictions.

\subsection{Adapting LIME to Dual-Stream EEG Models}
\label{sec:lime-adaptation}

LIME is a model-agnostic technique that explains individual predictions by learning a local surrogate model, typically a sparse linear regressor, around each instance of interest. It perturbs the input and observes output changes to infer feature importances. The relevance of each input component is estimated based on its contribution to the local decision boundary \cite{LIME16}.

To interpret the predictions of the bi-hemispheric model, we extend the LIME framework to support structured, dual-input neural architectures. The model under study receives EEG signals separately from the left and right hemispheres, which are then processed through independent branches. Since standard LIME is designed for flat, tabular input, we adapt it to handle this two-stream input format while preserving the spatial structure of the EEG signals.

Let $\Phi \in \mathbb{R}^{C \times F}$ be the frequency-domain EEG representation obtained after preprocessing and FFT, where $C$ denotes the number of channels, and $F$ is the number of frequency bins. The spectral data is split into two hemispheres:

\[
\Phi = \Phi^{\text{left}} \cup \Phi^{\text{right}}, \quad \Phi^{\text{left}}, \Phi^{\text{right}} \in \mathbb{R}^{C/2 \times F}
\]

Each hemisphere-specific input is processed through a sequence of operations: a 1D convolution, followed by max-pooling, dropout, and a recurrent (LSTM) layer. Let $f_{\text{conv}}, f_{\text{pool}}, f_{\text{drop}}, f_{\text{lstm}}$ represent these operations. The latent representations from each branch are:

\begin{align}
\mathbf{h}_L &= f_{\text{lstm}}(f_{\text{drop}}(f_{\text{pool}}(f_{\text{conv}}(\Phi^{\text{left}})))) \in \mathbb{R}^{H} \\
\mathbf{h}_R &= f_{\text{lstm}}(f_{\text{drop}}(f_{\text{pool}}(f_{\text{conv}}(\Phi^{\text{right}})))) \in \mathbb{R}^{H}
\end{align}

These vectors are concatenated as $\mathbf{h} = [\mathbf{h}_L; \mathbf{h}_R] \in \mathbb{R}^{2H}$ and passed through a dense layer and softmax to predict one of $K=6$ emotion classes.

\textbf{Flattening and Concatenating Inputs.}  
Each EEG sample consists of two matrices, one for each hemisphere. To make the input compatible with LIME, we reshape both matrices into 1D vectors and concatenate them, producing a single flattened vector $\mathbf{x} \in \mathbb{R}^{C \cdot F}$ that includes all channels and frequency features.

\textbf{Custom Prediction Function.}  
To ensure that LIME can still query the original model correctly, we define a custom prediction function that reverses the process of flattening. Let $\mathbf{x}_L \in \mathbb{R}^{C_L \cdot F}$ and $\mathbf{x}_R \in \mathbb{R}^{C_R \cdot F}$ represent the flattened left and right hemisphere features. The full input is $\mathbf{x} = [\mathbf{x}_L ; \mathbf{x}_R]$. We define:

\[
f'(\mathbf{x}) = f\left(\text{reshape}_L(\mathbf{x}_L), \text{reshape}_R(\mathbf{x}_R)\right)
\]

where $\text{reshape}_L$ and $\text{reshape}_R$ restore the original 2D format required by the model. This step is essential for maintaining architectural consistency during perturbation-based explanation.

\textbf{Generating Explanations.}  
For each sample, we apply LIME using the custom predictor and generate feature importance scores specific to the predicted emotion class. Perturbations are sampled around the original input, and their impact on the model's output is used to fit a local, interpretable model.

\textbf{Mapping Importances Back to EEG Channels.}  
Once local feature importances are computed for the flattened input, we re-map these values to their original EEG channels. We calculate the average importance score across frequency bins to obtain a single relevance score per channel. This is done separately for the left and right hemispheres, yielding a per-channel relevance profile that aligns with the bi-hemispheric structure.

Finally, we apply this procedure across all validation samples and aggregate the results. For each instance, we record the predicted class, ground truth label, class probabilities, and the corresponding LIME explanation. This adaptation enables LIME to provide interpretable, channel-level explanations for dual-stream EEG models, preserving the spatial organization of brain signals while exposing the internal reasoning of the classifier.

\section{Experiments and Results}
\subsection{Experimental Setup}

\begin{figure*}[t]  
    \centering
    \includegraphics[scale=0.3]{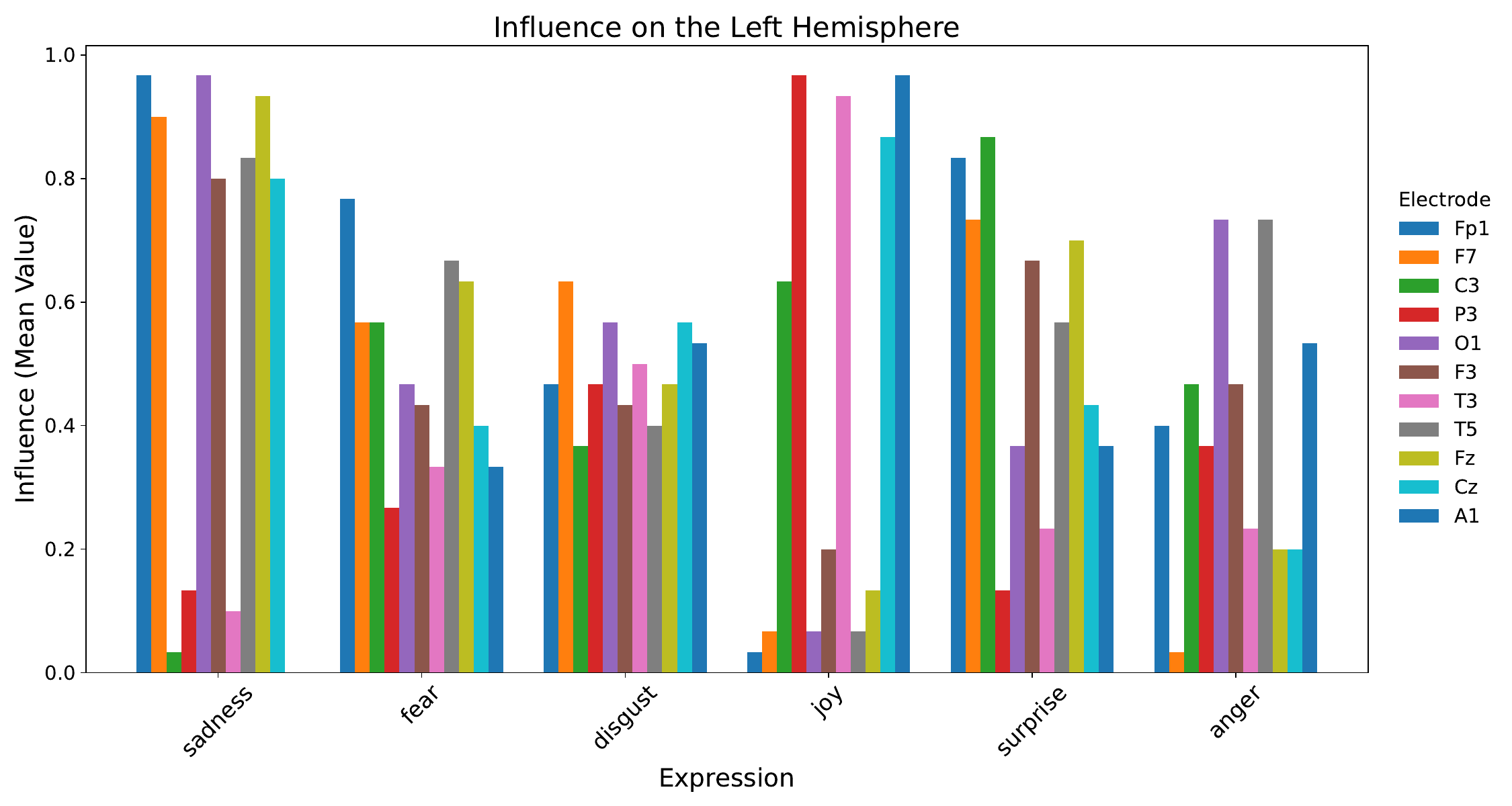}
    \caption{Influence of the left hemisphere electrodes on each emotion.}
     \label{fig:lhemisphere_influences}
\end{figure*}

\textbf{Dataset.}  
We conduct our experiments using the EmoNeuroDB dataset \cite{DubielFG24}, a multimodal corpus designed to support EEG-based emotion recognition in immersive virtual reality environments. The dataset comprises recordings from 40 participants (balanced by gender) who engaged in a VR-based emotional mimicry task. Each subject was exposed to an avatar displaying one of six fundamental emotions (joy, sadness, anger, fear, disgust, and surprise) and instructed to imitate the expression while wearing an EEG headset. Recordings were made with the DSI-24 wireless EEG system, using 21 dry electrodes arranged according to the 10-20 International System. Each emotion was repeated three times per subject, with each trial lasting approximately 15 seconds at a sampling rate of 300 Hz. Signals were preprocessed with band-pass filtering (1–50 Hz), re-referencing to mastoid electrodes, and correction for filter delay. The dataset is divided into training, validation, and test subsets. The training set contains data from 20 participants, while the validation set includes 10 participants. The remaining samples form an unlabeled test set used by the challenge organizers. Each EEG instance corresponds to one emotional expression and includes spectral features from 21 channels.

\textbf{Parameters and Evaluation Metrics.} 
The bi-hemispheric model is trained using the Adam optimizer and categorical cross-entropy loss. Input data are fed as frequency-domain features into two parallel branches (left and right hemispheres). Following the training setup described by the model's original authors, we use a batch size of 64, train for up to 200 epochs, and apply early stopping with a learning rate schedule. For LIME-based analysis, 5K perturbed samples are used per explanation, and channel-level relevance scores are aggregated across validation samples for group-level interpretation. Model performance is primarily assessed using classification accuracy on the validation set. To facilitate a deeper analysis, we also compute confusion matrices and class-wise performance metrics to examine the model’s ability to distinguish between emotions. These metrics provide insight into potential asymmetries or misclassifications relevant to emotion decoding.

\begin{figure*}[t]  
    \centering
    \includegraphics[scale=0.3]{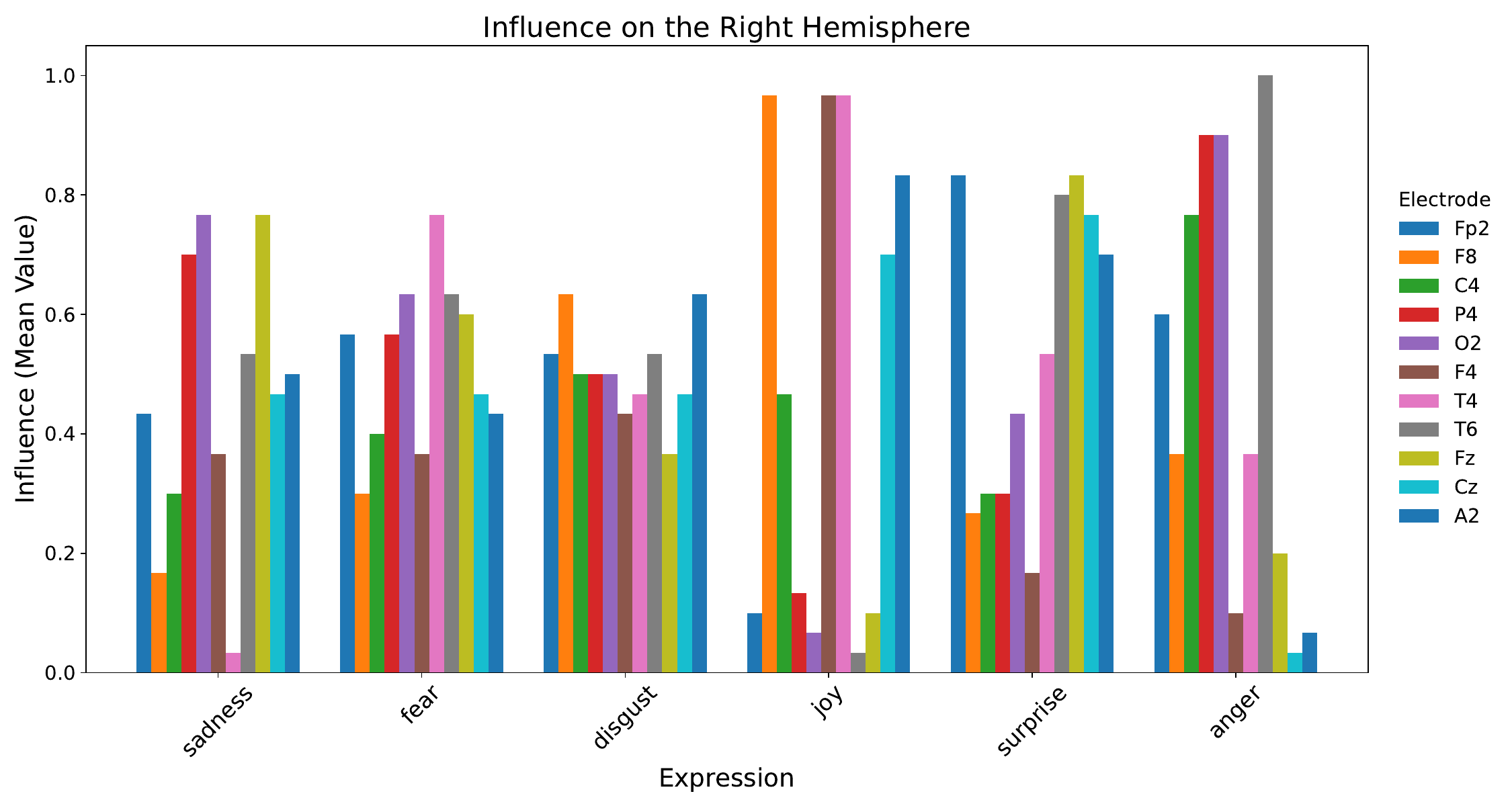}
    \caption{Influence of the right hemisphere electrodes on each emotion.}
     \label{fig:rhemisphere_influences}
\end{figure*}

\subsection{Local Explanations per Hemisphere}

To gain insight into the spatial relevance of the EEG signals per emotional class, we analyzed the average influence of each electrode on the left and right hemispheres separately. Figures~\ref{fig:lhemisphere_influences} and~\ref{fig:rhemisphere_influences} illustrate the contribution of each electrode, computed via local surrogate explanations, for every emotion in the validation dataset.

On the \textbf{left hemisphere} (Figure~\ref{fig:lhemisphere_influences}), certain frontal and central electrodes show consistently high influence. Notably, \textit{Fp1}, \textit{F3}, and \textit{Fz} exhibit high relevance for multiple emotions. For \textit{sadness}, electrodes \textit{Fp1}, \textit{O1}, \textit{F3}, \textit{T5}, and \textit{Fz} stand out with influence values above 0.8. In contrast, \textit{joy} shows a more distributed pattern, with peaks at \textit{P3}, \textit{T3}, \textit{Cz}, and \textit{A1}. \textit{Fear} and \textit{disgust} display moderate influence in temporal and parietal regions such as \textit{T3}, \textit{P3}, and \textit{T5}. Interestingly, \textit{Fz} remains influential across most emotional states, suggesting its central role in emotion decoding from the left side.

On the \textbf{right hemisphere} (Figure~\ref{fig:rhemisphere_influences}), the influence pattern varies more strongly across emotions. For example, during \textit{joy}, electrodes \textit{F8}, \textit{F4}, and \textit{T4} show dominant influence (above 0.9), suggesting right-lateralized activation in this emotional state. In contrast, \textit{anger} is characterized by maximal influence in posterior and temporal electrodes like \textit{T6}, \textit{P4}, and \textit{O2}. The central electrodes \textit{Cz} and \textit{Fz} also contribute significantly across emotions, although with more variability compared to their left counterparts. \textit{Fp2} and \textit{F8} are especially important for \textit{surprise} and \textit{fear}.

Overall, these local explanation maps reveal how different brain regions contribute to emotion classification and highlight notable asymmetries between hemispheres. For instance, emotional states such as \textit{joy} and \textit{anger} activate distinct electrode patterns across the two hemispheres, underscoring the importance of spatial decoding in EEG-based emotion recognition.

\begin{figure*}[t]  
    \centering
    \includegraphics[scale=0.35]{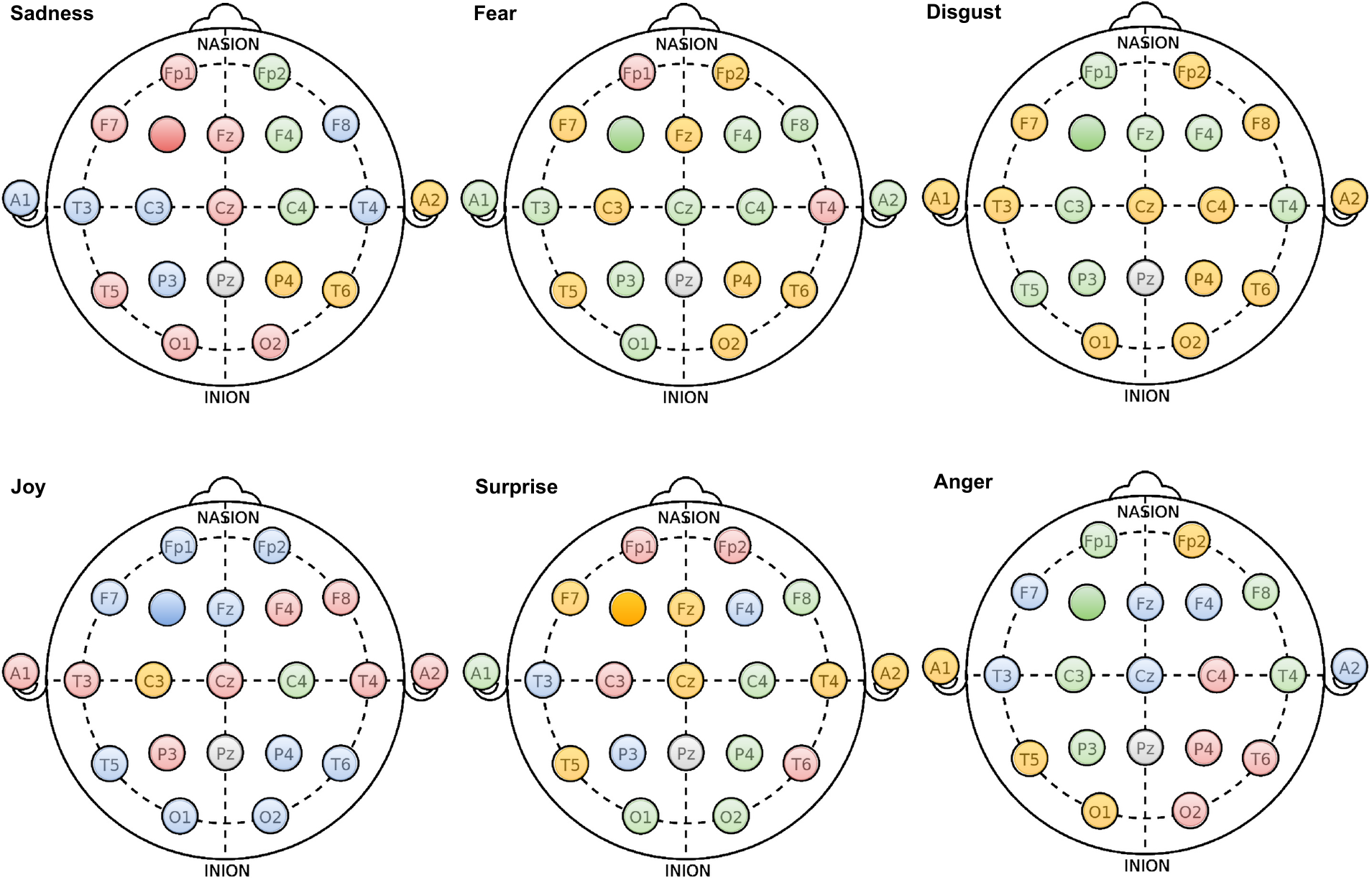}
    \caption{Topographic relevance maps. Red indicates very high relevance ($\geq$ 0.75), orange indicates high relevance (0.50–0.74), green indicates moderate relevance (0.25–0.49), and blue represents low relevance ($<$ 0.25). According to \cite{FreireFG24}, the \textit{Pz} electrode is used as reference location.}
     \label{fig:Topo}
\end{figure*}

\subsection{Topographic Relevance Maps}

To further interpret the spatial dynamics of emotional processing in EEG signals, we generated topographic relevance maps based on local explanation scores per electrode. These maps visualize the most influential brain regions across both hemispheres for each emotion, using averaged local surrogate explanations over the validation set. 

The relevance maps, shown in Figure~\ref{fig:Topo}, clearly highlight distinct spatial activation patterns associated with different emotional states. Each electrode on the topographic maps is color-coded based on its average influence score: red indicates very high relevance ($\geq$ 0.75), orange indicates high relevance (0.50–0.74), green indicates moderate relevance (0.25–0.49), and blue represents low relevance ($<$ 0.25). For example, \textit{sadness} and \textit{fear} exhibit strong influence in frontal and occipital regions of the left hemisphere, notably around electrodes \textit{Fp1}, \textit{F7}, and \textit{O1}. In contrast, \textit{anger} and \textit{joy} show stronger activations on the right side, particularly in temporal and parietal areas such as \textit{T6}, \textit{P4}, and \textit{F8}.

Interestingly, the central electrode \textit{Fz} demonstrates consistent relevance across most emotions, suggesting its role as a bilateral integrator in emotional processing. The lateral asymmetry observed, with emotions like \textit{disgust} showing more balanced activation, while \textit{surprise} is left-dominant, reinforces the importance of hemispheric specialization in EEG-based emotion decoding. These topographic insights offer a valuable perspective on how different brain regions contribute to emotion classification and validate the spatial patterns observed in the left and right hemisphere bar plots.

\subsection{Symmetric Electrode Correlations Across Emotions}

To explore the degree of functional symmetry between brain hemispheres under different emotional states, we computed the Pearson correlation coefficients between pairs of symmetric electrodes (e.g., \textit{Fp1–Fp2}, \textit{F3–F4}) across all instances labeled with the same emotion. These correlations were based on binarized importance values derived from the LIME explanations, where activations were thresholded and encoded as binary indicators of relevance.

\begin{figure*}[t]  
    \centering
    \includegraphics[scale=0.5]{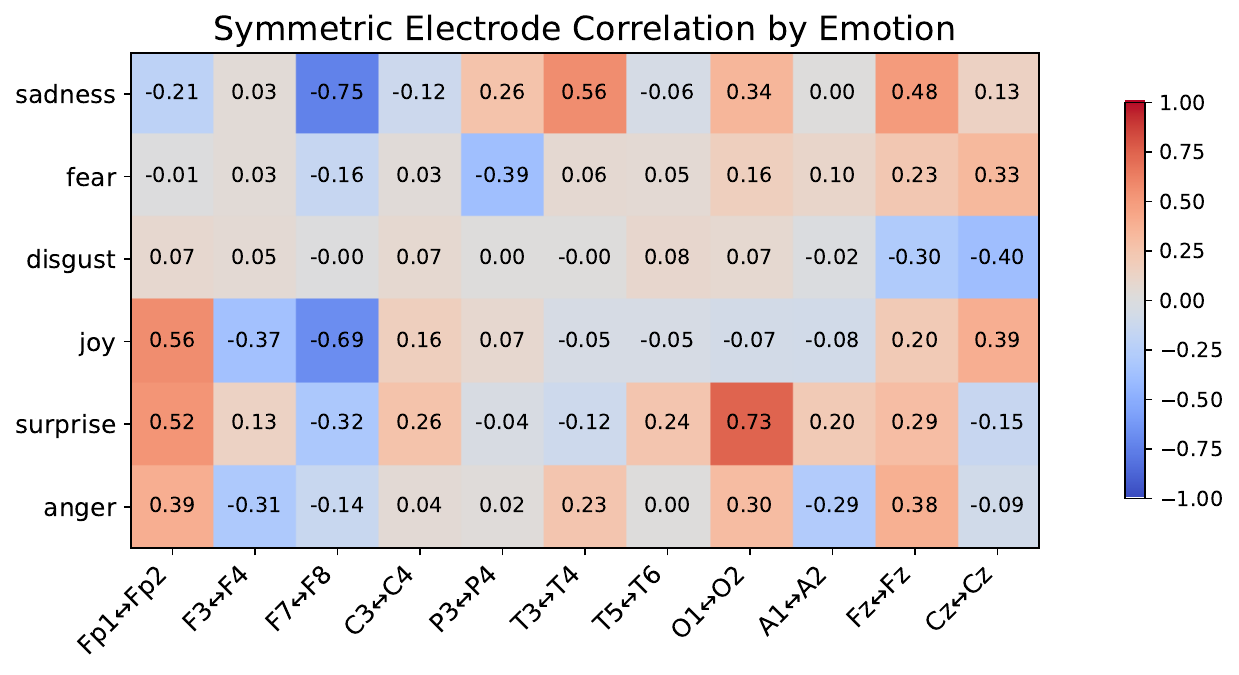}
    \caption{Average correlation of binarized relevance scores between symmetric electrode pairs for each emotion.}
     \label{fig:symmetric_heatmap}
\end{figure*}

Figure~\ref{fig:symmetric_heatmap} shows a heatmap depicting the average correlation values for each emotion across the 11 predefined symmetric electrode pairs. Higher correlation values indicate stronger bilateral engagement, while lower or negative values may suggest hemispheric asymmetries in how the model attends to features during classification. The color gradient in the heatmap spans from \textit{blue} (negative correlation) to \textit{red} (strong positive correlation), highlighting emotion-specific patterns of inter-hemispheric activation. Notably, emotions such as \textit{joy} and \textit{surprise} exhibited higher bilateral symmetry in frontal and parietal regions, while \textit{fear} and \textit{disgust} revealed more lateralized activation profiles.

The correlation analysis between symmetric electrode pairs provides insight into the lateralization of brain activity patterns during emotion classification. Overall, the results indicate variable degrees of hemispheric symmetry depending on the emotional category, suggesting distinct neural activation strategies adopted by the model across affective states.

Emotions such as \textit{joy} and \textit{surprise} consistently showed high positive correlations across multiple symmetric pairs, particularly in frontal (\textit{Fp1–Fp2}, \textit{F3–F4}) and parietal (\textit{P3–P4}) regions, implying that these affective states may elicit more balanced bilateral processing. This aligns with existing neurophysiological evidence suggesting that positive emotions tend to involve more symmetric or bilateral engagement of cortical areas, especially in non-pathological populations.

In contrast, emotions like \textit{fear} and \textit{disgust} demonstrated lower or even negative correlations in several symmetric pairs, indicating more asymmetric activation patterns. For instance, \textit{fear} displayed weak correlations in posterior and temporal regions, suggesting a potential lateralization effect, possibly reflecting right-hemisphere dominance often reported in affective neuroscience literature for threat-related stimuli.

Bearing in mind that, according to Figure \ref{fig:classifier}, both hemispheres include the central region, \textit{sadness} exhibited a mixed profile, with moderate symmetry in central regions (e.g., \textit{Cz–Cz}, \textit{Fz–Fz}) but reduced correlation in occipital areas, which may suggest more localized or unbalanced visual processing during the classification of this emotion.

These findings reflect that the model does not rely on uniform patterns of bilateral activation across emotions. Instead, its decision-making process appears to be emotion-specific, recruiting lateralized or symmetric cortical features depending on the affective state being classified. This behavior aligns with known functional asymmetries in the brain’s emotional circuitry, reinforcing the importance of considering hemispheric dynamics in affective computing systems.

\section{Conclusions}
\label{sec:conclusion}

This work introduces an explainable AI framework for interpreting bi-hemispheric EEG-based emotion recognition models by extending LIME to structured dual-input networks. The approach yields per-channel relevance maps that reveal how affective brain signals are processed across hemispheres.

Our analysis uncovers consistent emotion-specific hemispheric patterns: left frontal and central regions are more influential for \textit{sadness} and \textit{fear}, while right-hemispheric activity dominates for \textit{joy}, \textit{anger}, and \textit{disgust}. Relevance maps offer clear visualizations of these dynamics. Correlation analyses between symmetric electrode pairs show higher inter-hemispheric synchrony for positive emotions (e.g., \textit{joy}, \textit{surprise}) and more lateralized patterns for negative ones (e.g., \textit{fear}, \textit{disgust}), aligning with prior neuroscience findings \cite{davidson1992anterior}.

These findings demonstrate that deep learning models trained on EEG signals can capture meaningful spatial and lateralized patterns associated with emotional processing. The proposed framework bridges the gap between model accuracy and neuroscientific insight by introducing a tailored LIME extension for dual-input networks, enabling spatially grounded, emotion-specific explanations that are statistically robust and generalizable across subjects.

\textbf{Acknowledgments}.
This work is partially funded funded by project PID2021-122402OB-C22/MICIU/AEI
/10.13039/501100011033 FEDER, UE and by the ACIISI-Gobierno de Canarias and European FEDER funds under project ULPGC Facilities Net and Grant \mbox{EIS 2021 04}.

%
%
\bibliographystyle{splncs04}

\end{document}